\documentclass{article}

\usepackage{arxiv}
\usepackage[square,numbers]{natbib}
\usepackage[utf8x]{inputenc} 
\usepackage[T1]{fontenc}    
\usepackage{hyperref}       
\usepackage{url}            

\usepackage{booktabs}       
\usepackage{amsfonts}       
\usepackage{nicefrac}       
\usepackage{microtype}      
\usepackage{lipsum}
\usepackage{authblk}
\usepackage{tabularx}
\usepackage{graphicx}
\usepackage{caption}
\usepackage{setspace}

\title{Applications of Blockchain in Healthcare: Current Landscape \& Challenges}
\author[1]{Gajendra J. Katuwal}
\author[2]{Sandip Pandey}
\author[3]{Mark Hennessey}
\author[3]{Bishal Lamichhane}

\affil[1]{Philips Research North America, Cambridge, MA, United States}
\affil[2]{Blockchain Lab, Delft University of Technology, Delft, Netherlands}
\affil[3]{Philips Research Europe, Eindhoven, Netherlands}

\begin{document}
\maketitle

\let\thefootnote\relax\footnote{

 \linespread{1} 
\tiny \textbf{Disclaimer:} This work is not a systematic review of the field, is purely academic in the intention and is no way biased by the affiliation of the authors. The primary sources of information for our review work are the company webpages, healthcare news portal, and academic journals. Application of blockchain in the field of healthcare is a new and growing field. Therefore, a lot of information made available might be aspirations and speculations, not necessarily translating to mature concepts, products, or services. We have given the care to include credible works in this review. However, fully confirming the credibility of the provided information, e.g. checking if a blockchain enabled new product/service offering matches the claims made, is beyond the scope of this review work. The work mentioned in this paper should be assessed accordingly by the readers. Further, it is not possible to include all the companies and initiatives active in the field due to the mere number of such initiatives. The companies and initiatives enlisted in this paper have been included just based on the ease of assessing the information from the given company or initiative. This review paper is in no way endorsing or promoting the products and offerings from the enlisted companies. Further, our review here is also not a statement of the merit of work for the enlisted companies compared to other works not included in the review. }

\begin{abstract}
 Several problems in healthcare stem from the complex network of intermediaries and the lack of traceability of transactions. To mention a few: healthcare data is fragmented across several silos negatively affecting research and services, about half of the clinical trials are never reported, the cost of drug discovery is ever increasing, and substandard and fake medicines are still a huge problem. Blockchain has the potential to solve these problems as it provides trust without any intermediaries, has traceability as a default feature, and promises new business models by enabling novel incentive structures. Due to its potential, blockchain has gathered significant interest in the healthcare industry. In this paper, we review major use cases of blockchain in healthcare: patient data management, pharmaceutical research, supply chain management of medical goods, prescription management, billing claims management, analytics, and telemedicine alongside the related projects. We found that most of the blockchain projects are limited as white-papers, proof of concepts, and products with a limited user base. However, we observed that the quantity, quality, and maturity of the projects are increasing. We also discuss technical, regulatory, and business challenges to the adoption of blockchain in the healthcare industry. 

\end{abstract}

\keywords{Blockchain \and Healthcare \and Trust  \and  Traceability \and Data Management \and Supply Chain \and Analytics}

\section{Introduction}
The healthcare industry is one of the world's largest industries, consuming over 10\% of gross domestic product (GDP) of the most developed nations\cite{health_spending}. Simply put, this industry includes generalization and commercialization of goods and services to treat patients with curative, preventive, rehabilitative, and palliative care. Being a complex system of interconnected entities under heavy regulatory boundaries, patient data is highly fragmented and the cost of healthcare delivery is continuously rising due to inefficiencies in the system and dependence on several intermediaries. Furthermore, transparency on the whole process of enabling data sharing between multiple parties, even though supposedly beneficial to the patient, is still lacking on full transparency and control from the patient's view. Patients have shown concern about the possibility of their data being used by for-profit entities \cite{bell2014sharing}. This has accentuated a need for an information technology system that can remove the middlemen and cut costs while maintaining trust and transparency. The blockchain is a revolutionary technology which can assist in solving the challenges of healthcare by providing decentralized trust. Blockchain-enabled decentralization promises to minimize the problem of vendor lock-in that has plagued the healthcare industry.

Patient data is scattered across different entities in the value chain of the healthcare industry referred to as data silos and sharing of data is prone to a multi-level process of permission control. Because of this, oftentimes crucial data is not accessible and available at the time of urgent need. Blockchain can solve this issue with health information exchange (HIE) by serving as a basis for a trusted decentralized database. It can enable one-stop access to the entire medical history of a patient across all healthcare providers. Access control system built using trust on blockchain puts patients in control of their data; they can grant consent and access rights to external parties like researchers to have access to all or subset of their medical records. This feature fits nicely with the patient-centric model of healthcare where blockchain can act as a catalyst inducing trust.

The records written on the blockchain are immutable and cannot be altered or deleted. This characteristic of blockchain provides primitives like data integrity and provenance which can be used to build solutions to prevent drug counterfeiting and medical frauds. For instance, fraudulent results and removal of data in clinical trials which do not align with researcher's bias or funding source can be prevented by enforcing the integrity of data in blockchain. In addition, it allows keeping an immutable log of subject's consent in a clinical trial. On a financial note, blockchain could save hundreds of billions for the pharmaceutical industry by defining a chain-of-custody in the supply chain\cite{pwc_counterfeit}.

It is possible to write custom laws and rules forming contracts on the blockchain, which are equivalent to real-world contracts and can be legally binding. These contracts are referred to as smart contracts. Smart contracts can be used in several processes within healthcare including billing and insurance which helps in automating the process and reducing the costs. Later in subsequent sections, we see several companies leveraging the smart contracts to build solutions.

In this publication, we discuss the problems in the healthcare ecosystem and how the right use of blockchain technology may solve these problems. We begin by giving a brief introduction to inner workings of blockchain in Chapter \ref{sec:blockchain}. In subsequent chapters, we review major use cases of blockchain in healthcare along with the projects pursuing these use cases. We have reviewed works in the field of healthcare data management, pharmaceutical use cases like supply chain and medication adherence, billing/claims management, analytics, telemedicine, and blockchain as a service. These fields, though not all-encompassing, represent some of the major areas in the overall healthcare space. In Chapter \ref{sec:challenges}, we discuss the challenges in the adoption of blockchain based solutions in healthcare. The final chapter concludes with the overall discussion on the landscape of healthcare, problems therein, promises of the blockchain, and the state of blockchain based solutions. 

Several other review papers have discussed the applications of blockchain in healthcare \cite{Kuo2017, Dhillon2017, Engelhardt2017, Randall2017, Sharma2018_FCCCO}.  

\section{Blockchain} \label{sec:blockchain}
Blockchain is the technology behind Bitcoin \cite{nakamoto2008bitcoin}, an open peer-to-peer (p2p) value transfer network that solved the problem of \textit{double spending} for the first time.
With the release of Bitcoin, cryptocurrencies became the first use case of the blockchain. Cryptocurrencies are analogous to fiat money like USD or EUR facilitating the exchange of value but use cryptographic protocols as the basis for governance rather than depending on a central authority like banks. In recent years, there has been increasing awareness that the blockchain has much more to offer than cryptocurrencies; blockchain can be an ideal tool for creating \textit{trust based solutions}.

In Bitcoin like cryptocurrencies, a blockchain represents \textit{an append-only linear chain of blocks or ledger replicated across a network of computers}; a block is a group of validated transactions in the network, and the two consequent blocks are cryptographically linked. However, in recent times we observe variations on how blockchain is defined. Often it is used as an umbrella term for the distributed ledger technology used in any  p2p network that tries to solve problems like double spending. In this publication, we broadly consider blockchain as a p2p distributed append-only ledger with immutable and tamper-proof records.  It is important to understand the concept and life cycle of a transaction to understand how blockchain works. In blockchain, a transaction represents a change of state. 
For instance, a transfer of coins from a sender to a receiver is a transaction where the account balance represents the state. When a new transaction is created, it is broadcasted to the network where a miner node (computer) picks up the transaction and compose a block by combining one or more transactions and broadcasts the block into the network. Each blockchain has a set of rules that determine whether a block created by a miner is valid or not, and only the valid blocks are added to the chain. For example, a block is considered valid and mined in Bitcoin if it has the correct cryptographic hash \cite{rogaway2004cryptographic}. These set of rules form the basis of consensus in the blockchain.

\subsection*{Consensus}
The consensus in a blockchain network is the mechanism on agreeing on a common version or state of the blockchain which is considered to be the trusted truth of the blockchain. The consensus mechanism prevents the \textit{double spending} as trying to double spend an already spent transaction causes a conflict with the existing state, thereby the transaction is rejected and will never be added to the chain. A consensus protocol includes the rules for transaction validation, accepting the newly created block into the chain, and selection of fork/partition in case of network partitions. Depending on the context and the use case, the need and requirement for the consensus could differ. Consensus mechanism can be broadly classified as global or local.
\begin{enumerate}
    \item In \textit{global consensus model}, the first block of the chain called genesis block is common for all nodes in the network and every node agrees on the same state of the network and stores the complete chain to validate any transaction. Most common examples of global consensus blockchain are Bitcoin and Ethereum.
    \item In \textit{local consensus model}, every participant owns an individual genesis block and the consensus is only reached among the parties involved in the transactions. This local consensus reduces the storage requirements on individual nodes and is generally more scalable than global consensus counterparts. Example blockchains are TrustChain \cite{otte2017trustchain} and Nano \cite{nano_whitepaper}.
\end{enumerate}

Most of the current blockchains use some version of the consensus protocol of Bitcoin called  Nakamoto consensus \cite{nakamoto2008bitcoin}. The Nakamato consensus is a global consensus model; it uses proof of work (PoW) for accepting the new block and dictates to choose the longest chain in case of network partitions. In PoW, miners compete to find a cryptographic hash that solves a computationally difficult problem and the block proposed by the miner with the correct hash is added to the chain. Proof of work is resource intensive and inherently non-scalable. Proof of stake (PoS) is an alternative block validation model where the creator of the new block is chosen in a deterministic way depending on its wealth, also called the stake \cite{saleh2018blockchain}. PoS is much more efficient in terms of performance and energy consumption. We are observing a shift from PoW to PoS or hybrid model in recent blockchains where Ethereum remains a strong proponent.

\subsection*{Smart Contracts}
A smart contract represents a self-executing, self-verifying, and tamper-resistant piece of code with programmable application logic which resides and executes on the blockchain. It formalizes the transaction rules and relationship among entities and assets in the blockchain and gives the flexibility to write custom application logic which becomes a law enforced by the blockchain itself without any reliance on trusted intermediaries. Smart contracts form the basis for trust in the application layer.

From the early days of Bitcoin, it supported a basic scripting language \cite{bitcoinscript} that allowed people to write custom logic to spend transactions. This language gave the people the flexibility to design primitive contracts. For example, enforcing a transaction to spend only after a certain time or requiring a multi-party signature to spend a transaction became possible. 
Though Bitcoin scripts provided a way to write contracts, it was limited and restrictive. Then, the blockchains specifically designed to support the execution of custom code came into existence. Ethereum \cite{wood2014ethereum} is a pioneer blockchain explicitly designed as the smart contracts platform. With smart contracts, real-world rules and laws can be embedded and enforced into the blockchain, and thus developing a  decentralized trusted application (DApp) became a lot easier. Cryptocurrency token and digital assets became the most popular contract applications. Soon an outburst of DApps was seen contributing to a world-wide blockchain hype. Prediction markets \cite{Peterson2018_augur_whitepaper} and decentralized exchanges \cite{zrx_whitepaper} are some examples which demonstrate the real strength and flexibility of smart contracts.

 Apart from Ethereum, there are several other blockchain platforms supporting smart contracts: NEM \cite{nem}, Hyperledger Fabric \cite{androulaki2018hyperledger}, Stellar \cite{stellar}, NEO \cite{neo-project}, etc.

\subsection*{Types of blockchain}
Blockchain can be public or private depending on the permission level \cite{zheng2017overview}, but here we broadly classify it to three different levels of granularity.
\begin{itemize}
\item \textbf{Public blockchain}
A public blockchain is permissionless, and anyone can easily participate and validate the transactions. Transactions are public and anonymous/pseudonymous. The blockchain is maintained by the public community, so there is the highest level of decentralized trust. Bitcoin is the pioneer public blockchain. Bitcoin, Ethereum, Waves \cite{Waves_whitepaper}, Dash \cite{dash_whitepaper}, and Bitshares \cite{bts_whitepaper} are few examples of public blockchains. 

\item \textbf{Federated blockchain}
A federated blockchain is a permissioned blockchain operating under the leadership of a group often called the consortium. Predefined consortium nodes control the consensus. The transactions may or may not be public. Some examples includes R3 Corda \cite{brown2017introducing}, EWF (Energy) \cite{EnergyWebF}, and B3i (Insurance) \cite{B3iInsurance}. 

\item \textbf{Private blockchain}
A private blockchain is a permissioned blockchain centralized to one governing organization. Transactions are validated internally and may or may not be public readable. Private blockchains usually have faster block times and can process higher transaction throughput. However, these are vulnerable to security breaches. The value of private blockchain can be seen as a trust transformer where trust is based on an algorithm rather than an authority. Monax \cite{MonaxBlockchain}, HyperLedger with Sawtooth \cite{Sawtooth}, private Ethereum are a few examples of private blockchains. 
\end{itemize}

\subsection*{Why blockchain?}
Trust and traceability are the two basic promises of the blockchain obtained out of the box which solves the generic trust problem on all public, federated, and organization levels. However, these traits are not always sufficient to provide a complete solution, which is why we often see blockchain paired with strong cryptographic protocols like zero-knowledge proofs \cite{goldreich1994definitions}. 
This pairing ensures to provide trust, traceability, security, and control which are the core building blocks for critical solutions in several industries including health care and supply chain.
Data recorded in the blockchain cannot be changed or deleted without leaving a trace. This immutability and traceability of the data is a critical requirement for any health care system. Thus, the benefits of blockchain seem imminent. Here, we list some of the core issues/concerns that need to be addressed in health care solutions and later show how blockchain could help in solving them. 
\begin{itemize}
    \item Secure storage and integrity protection
    \item Privacy and ownership of data
    \item Data sharing
    \item Traceability and accountability of data
\end{itemize}
While each of these concerns can be addressed separately with the proper use of cryptography and privacy-preserving technologies, the key concerns in such solutions have always been the governing trust model. In such solutions, blockchain as a trustful decentralized ledger technology can act as a trust binding glue. 

\subsection*{Challenges \& Future}
Despite the massive potential, there are limitations as of the current state of the blockchain. Currently, every node in the network processes the transaction which makes the blockchain rather slow and unsuitable to handle real-world transactions which range in tens of thousands of transactions per second. This disparity highlights the scaling issue that blockchains have to overcome for wider adoption across all industries. Moreover, with the growth in usage, the size of blockchain is increasing enormously, making it difficult for normal users to keep the full copy of it. On a positive note, with huge investment and research efforts invested into blockchain \cite{rowley_2018}, a better scalable blockchain may evolve in the future.

Decentralization, consistency, and scalability (DCS) are the three desirable properties that blockchain platforms strive to maintain. A blockchain system can only simultaneously provide two out of three properties; there will always be a trade-off. Platforms like Bitcoin and Ethereum are DC system where they provide sufficient decentralization and consistency of data but lack scalability. On the other hand, Hyperledger is an example of a CS system which ensures data consistency and can scale above 10K+ transactions per second at the cost of losing decentralization.

In the last decade, we have observed the evolution of blockchain in three generations. Cryptocurrencies represented the first 1.0 generation which was primarily designed as an alternative payment system. Then, decentralized applications based on smart contracts (Dapps) represented the 2.0 applications providing business logic abstraction and execution on a trusted platform where safety and security of smart contracts is the key issue. Smart contracts which can be validated and tested before deployment on a live blockchain prevents financial losses due to flaws in the code. The third generation is taking a pervasive multi-dimensional approach connecting Internet of Things (IoT), machine learning, and different branches of science. These different generations of blockchain are evolving simultaneously in their own pace, addressing the issues like scalability, security, and privacy along the way. Moreover, projects like Holochain \cite{holochain_whitepaper} and Hashgraph \cite{Hydera-hashgraph_whitepaper} are developing scalable and general purpose platforms for agent-centric decentralized applications; traditional blockchains are data-centric making them difficult to scale. We see more of these example use cases in the following sections.

\section{Healthcare Data Management} \label{sec:data_management}
The management of healthcare data which includes storage, access control, and sharing of the data is an important aspect of the healthcare industry. Proper management of healthcare data improves healthcare outcomes by allowing holistic views of patients, personalized treatments, and efficient communication. It is also critical for operating healthcare industry cost-effectively and efficiently. However, managing healthcare data is a challenging task due to its sensitive nature and subsequent trust issues. And it is one of the main reasons why the healthcare system is disconnected—healthcare data and services exist in disparate forms in several silos. This disconnected system is a culprit for several inefficiencies in healthcare and is a major hurdle for healthcare research.  Healthcare professionals generally do not have access to the complete data of patients, thereby, hampering the subsequent diagnosis and treatment steps; and researchers struggle to find the desired data for their studies, thereby, slowing down healthcare research.

Blockchain may enable the efficient sharing of healthcare data while ensuring data integrity and protecting patient privacy. Secure, efficient, cost-effective, and interoperable HIE can be built with its right use alongside with other technologies. Moreover, the adoption of blockchain can push forward the movement of patient-centric healthcare model where patients control their healthcare data. The major hurdles behind data-sharing in both patient-centric and traditional models are \textit{lack of trust} and \textit{lack of incentives to share}. The blockchain technology can solve both problems by acting as a trust layer and introducing the incentive mechanisms such as rewarding crypto tokens for sharing data. Moreover, blockchain can be the bridge for the integration of medical device data and healthcare internet of things; the healthcare and lifestyle data collected by wearable devices can be critical for correct diagnosis but are underutilized since there is a lack of a proper way for a physician to access the patient-generated data. 

With \textit{blockchain-enabled trust and incentive structure}, there is a promise for a global HIE and a marketplace on top of it. But the lack of common healthcare data standards can be one of the major obstacles to overcome before the development of an interoperable HIE connecting multiple disparate data silos. However, there is a possibility that the incentives introduced by blockchain-based data exchange may fuel the creation and the development of the open data standards. Blockchain-based HIE will be an interesting use case that requires the balance among privacy, transparency, and efficiency.  Moreover, country-specific regulations will be another obstacle for a HIE connecting multiple regulatory regions. Since patients have full ownership of their healthcare data in many countries, blockchain enabled patient-centric healthcare data model can be one fitting way to bypass these regulatory challenges.

\subsection{Blockchain based projects for healthcare data management}
Several projects are focusing on developing some form of blockchain based HIE and providing data and services marketplace on top of it; see Table \ref{tab:data_management}. Among them, some are targeting general electronic health records (EHR) data while some are specializing in particular data modalities such as genomics and dermatology.
For example, Medrec \cite{Ekblaw2017MedRec} is an open source blockchain platform for EHR management. It was recently tested in collaboration with Beth Israel Deaconess Center.
Patientory \cite{Mcfarlane2017} is one of the early blockchain based healthcare startup leveraging Initial Coin Offerings (ICO) for funding.  It is developing a HIE powered by its own blockchain.
HealthSuite Insights of Philips Healthcare is testing Verifiable Data Exchange Process, a product that enables the secure and traceable data exchange between the members inside a network of hospitals and universities
 \cite{Philips_TNW_Oct2018}; all the data exchanges inside the network are stored in a blockchain alongside the identities of the people performing the exchanges to create an audit trail of the data exchange. Medshare \cite{Xia2017} provides a blockchain based data sharing of electronic medical records among untrusted parties by introducing data provenance, auditing, and trailing on medical data. ﻿Utilizing smart contracts and an access control system, they claim that their system can effectively trace the behavior of the data and revoke access to violated rules and permissions on the data.
Iryo \cite{Iryo_whitepaper} is creating a global repository of health data in OpenEHR format \cite{openEHR_website}. See Table \ref{tab:data_management} for other similar initiatives on the use of blockchain-based solutions for healthcare data management.

\begin{table}[]
\centering
\caption{Example initiatives on the use of blockchain for management of healthcare data \& services}
\label{tab:data_management}
\begin{tabularx}{\linewidth}{p{0.1 \linewidth}  p{0.2\linewidth} | p{0.6\linewidth}}

        \toprule
        \textbf{Group} & \textbf{Comments}  & \textbf{Projects}\\
        \midrule
        EHR &  & 
        Medrec \cite{Ekblaw2017MedRec}, 
        Patientory \cite{Mcfarlane2017}, 
        HealthSuite Insights Philips Healthcare \cite{Philips_TNW_Oct2018},
        Gem Health \cite{gem_health},
        Medshare \cite{Xia2017},
        Iryo \cite{Iryo_whitepaper},
        FHIR Chain \cite{Zhang2018_FHIRChain},
        OMNI PHR \cite{Roehrs2017_OMNIphr},
        Medicalchain \cite{medicalchain_whitepaper},
        Doc.ai \cite{doc.ai_website},
        Hearthy \cite{hearthy.co_whitepaper}
        \\ \\
        & Focus on developing countries &
        Factom \cite{factom_gates_healthcare}
        \\ \\
        Genomics &  & 
        Encrypgen \cite{encrypgen_website},
        Nebula Genomics \cite{nebula_genomics_whitepaper},
        lunaDNA \cite{luna_dna_website},
        Zenome \cite{zenome_whitepaper},
        Genomes.io \cite{hahnel2018_genomes_whitepaper},
        Shivom \cite{shivom_website}
        \\ \\
        Imaging &  & 

        ETDB-Caltech \cite{Ortega2018_ETDB-Caltech},
        Patel et. al 2018 \cite{Patel2018_image_exchange}
        \\
          & Dermatolgoy  & 
         OPU Labs\cite{OPU_whitepaper},
         MedX Protocol\cite{medxprotocol_medcredits_website},
         Dermonet \cite{Dermnet_italy_paper}
         \\ \\ 
         & Network as a service & 
         Akiri switch \cite{akiri}
       \\ \bottomrule
    \end{tabularx}
    
\end{table}



Multiple projects are focusing on particular modalities of data such as genomics and imaging. Genomics, in particular, has attracted a lot of interest from entrepreneurs and companies probably because of the recent popularity of personal genome sequencing, the importance of genomics data, and an immense possibility of its monetization. Personal genomics companies such as 23andMe and AncestryDNA monetize the genetic data by selling access to third parties such as labs and biotech companies. Several startups such as Encrypgen \cite{Encrypgen_whitepaper2017}, Nebula Genomics \cite{nebula_genomics_whitepaper}, LunaDNA \cite{luna_dna_website}, etc. are developing blockchain based genomics data-exchange platform or network. With the blockchain-based platforms, they claim to reduce the cost of genome sequencing, to give control of the data to patients, and to share the value captured from the monetization of the data to patients.

\section{Pharmaceutical sector}
\begin{figure}[!htb]
\centering
\includegraphics[width=\textwidth]{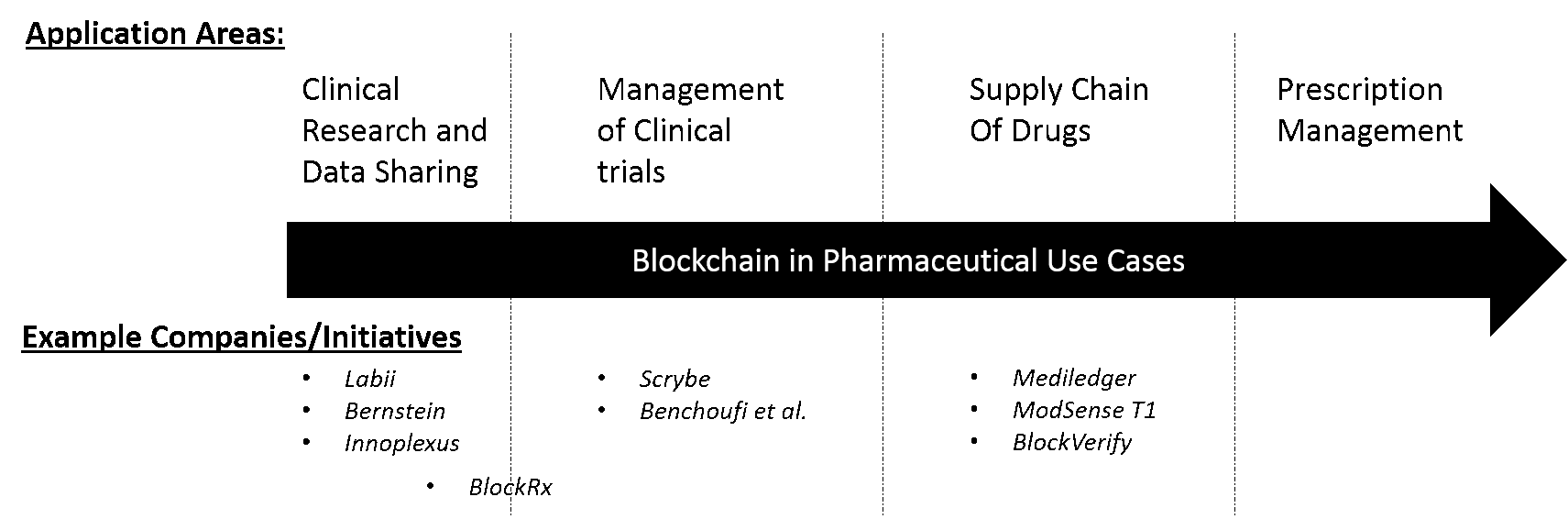}
\caption{Overview of the reviewed application areas for the use cases of blockchain in the pharmaceutical areas. Per application areas, examples of companies or initiatives involved in that particular area are also provided. }
\label{fig:fig_pharma_blockchain} 
\end{figure}

Pharmaceutical supplies are an important aspect of clinical care and healthcare delivery. In this section, we review the various innovative applications and initiatives in the pharmaceutical sector, covering the entire spectrum right from the drug discovery and clinical trials for the market introduction to solutions at the end of the chain like counterfeit drugs identification and patient adherence to medication. An overview of the reviewed application areas and companies within those application areas is shown in Figure \ref{fig:fig_pharma_blockchain}. 

\subsection{Drug discovery and Pharmaceutical research}
Drug discovery and research take a significant cost on the operations of any pharmaceutical company. With increasing costs of healthcare, together with the need to innovate faster on new medicinal treatments, it is imperative that multiple pharmaceutical companies find an approach to collaborate competitively. Blockchain can enable the technological platform to facilitate the transfer of trusted information and knowledge among multiple parties. The usage of blockchain for robust digital proof of Intellectual Property (IP) through immutable records and time-stamping is one fitting proposition for the collaboration. Blockchain-based solutions can also provide mechanisms to share clinical and trial data competitively. Even under a non-collaborative research and drug development scenario, blockchain provides benefits for effectively tracking and managing various aspects of clinical trials like data management, consent management, tracking side-effects of drugs usage, etc. Also, it is not uncommon for a pharmaceutical research company to outsource their clinical research projects. In this case, blockchain could provide a feasible mechanism to assure data integrity and proper outcome validation. In the current system, the pharmaceutical companies might have incentives to misrepresent results, e.g. in disclosing the side-effect of new drugs. With an open research ecosystem based on blockchain technology, research outputs are transparent, and research outputs are validated making the false representation of the results difficult. 

\textbf{IP management.} Many blockchain based solutions for IP management have been proposed in the generic space, which can also be applied for the innovations in drug development. An example, in this case, is the ambition to use blockchain backed electronic lab notebook solutions from Labii \cite{Labii2018}. Bernstein \cite{Berstein18} provides a blockchain based management of digital trails with timestamping to safeguard priorities in IP; this feature can be beneficial in collaborative pharmaceutical research. A solution from iPlexus \cite{iPlexus18} uses blockchain to make any unpublished and published data from drug development studies readily available for use. Blockchain solves the puzzle of maintaining trust and protecting IPs to enable such a groundbreaking initiative and framework.

\textbf{Clinical Trials.} Blockchain also has use cases in the management of the clinical trial process for pharmaceutical research. Recently, IEEE Standard Association \cite{ieee_blockchain_for_clinical_trials} organized a forum on Blockchain for Clinical Trial with the aim to use blockchain to make innovations in patient recruitment, ensure data integrity, and make rapid advances in drug development. Scrybe \cite{worley18}, a blockchain project presented in the forum, enables an efficient and trusted mechanism to expedite the clinical trials and research process. Of others, it allows an easy and transparent framework for legal and ethical validations of the trial process by the auditors. The work in \cite{Ravaud17} shows how blockchain can be used to manage the consent, data, and outcome from a clinical trial in a trustful and open manner. Such innovations in clinical trial rollout and management are crucial for advancements in pharmaceutical research. A lot of clinical trials run over the budget and timelines. Competitive sharing of clinical and trial data can accelerate research and discovery. Such sharing of data can be enabled with blockchain, using example approaches as discussed in Section \ref{sec:data_management}.

 Further, the research aspect of pharmaceuticals is quite broad that pervades the drug discovery process to device manufacturers and clinical trial outcomes. A solution across this spectrum is provided by BlockRX \cite{BlockRx17} using so-called advanced digital ledger technology (ADLT). The overarching goal is to inter-connect the currently disconnected parties in silos. 

\subsection{Supply chain and Counterfeit drugs detection}
The importance of the supply chain in the medical industry cannot be overstated. Right from the raw materials and production, to different stages of storage and distribution, proper monitoring and tracking are required to ensure optimal and intended usage. One of the growing concerns in recent years is that of counterfeit drugs. There must be a mechanism for the end users and all the stakeholders of the supply chain to verify the ingredients of a drug. With lack of a proper tracking mechanism, there are ample weak links in the supply chain where the drug can be tampered with, or a counterfeit drug can be infused in the supply. To address this growing concern, new regulations have been brought forth to have all stakeholders of the medicine supply chain provide a robust mechanism to track and trace the pharmaceutical supplies that pass through them. Blockchain provides a perfectly fitting solution to this need for tracing and tracking, where this information has to be maintained in an open yet safe and tamper-proof system accessible to multiple parties. 

Many solutions have therefore been proposed using blockchain to track the supply chain of pharmaceutical supplies. 
The MediLedger project \cite{MediLedger2017} is building an open network for the pharmaceutical supply chain. The network is a permissioned blockchain for the partners involved in the supply chain of pharmaceuticals. The concept is in line with the track and trace regulations. The law requires a way to track prescription drugs through the entire supply chain using interoperable systems, which MediLedger, a project launched by Pharma giants like Pfizer and Genentech, is proposing to solve using blockchain technology.  One of the most ambitious projects in this space is Ambrosus \cite{ambrosus_whitepaper} with its flagship product AMB-net, a blockchain based IoT network for supply chain targeted for food and pharmaceutical industries. On top of AMB-net, companies can build their custom supply chain solutions.
MODsense T1 from Modum \cite{Modum18} provides a temperature sensor for monitoring the conditions in the supply chain of pharmaceuticals, helping meet the regulatory requirements on Good Distribution Practice (GDP) of medicinal products for human use. The sensor data and other digital records are maintained using a blockchain. Several other blockchain based solutions in the supply chain industry also list the pharmaceutical industry as their primary application areas. BlockVerify \cite{Blockverify18}, for example, list pharmaceuticals as one of the main application areas for their supply chain solution preventing counterfeits. In the provided solution, the history of the goods is recorded with the BlockVerify tag on a blockchain.

An overview of some of the initiatives of blockchain enabled supply chain solutions for the pharmaceutical applications is given in Table \ref{tab:pharma_supplychain}.

\begin{table}[]
\caption{Example initiatives on the use of blockchain for pharmaceutical supply chain solutions}
\label{tab:pharma_supplychain}
\begin{tabularx}{\linewidth}{p{0.35\linewidth} | p{0.6\linewidth}}
\toprule
Initiatives & Description \\
\midrule
MediLedgerProject \cite{MediLedger2017}                                             & Permissioned blockchain solutions to meet the track and trace regulation in pharmaceutical supply chain.  
\\
Ambrosus 
\cite{ambrosus_whitepaper}            &
AMB-net, a blockchain based IoT network for supply chain targeted for food and pharmaceutical industries.
\\
Modsense T1 from Modum  \cite{Modum18}                                        & Blockchain based tracking of temperature and environment conditions along the supply chain.                                                                                                                            \\
Blockverify \cite{Blockverify18}                                                   & Anti-counterfeit and transparency solution for supply chain with applications to pharmaceutical sector among others.                                                                                                   \\
DHL collaboration with Accenture \cite{DHLAccenture}                                                         & The initiative, dubbed as prototype solution service, uses blockchain to track pharmaceutical products throughout the entire supply chain.                                                                             \\
Imperial Logistics collaboration with One Network Enterprises \cite{ImperialOneNetwork} & The solution from this collaboration is intended to improve supply chain security using the One Blockchain platform from the One Network Enterprises.                                                                  \\
Authentag  \cite{Authentag}                                                 & Provide distributed ledger technology from blockchain to provide tracking and verification services for pharmaceutical supply chain.                                                                                   \\
EasySight Supply chain management/ Hejia \cite{EasySight}                                                        & With the motivation to enable smaller company have reduced time for receiving payments, the blockchain based solution from EasySight tracks drugs through the supply chain for complete transparency of trade records. \\
GFT \cite{GFT} collaboration with MYTIGATE \cite{Mytigate}                  & The solution from this collaboration is a proof of concept on the use of blockchain to keep track of pharmaceuticals.                                                                                                  \\
SAP                                                                                       & SAP has combined blockchain technology with their ATTP (Advanced track and trace for pharmaceuticals) to solve supply chain issues considering new regulatory requirements.                                            \\
IEEE Pharma Supply BlockChain Forum \cite{IEEEPharma}    & A general podium for multiple stakeholders to initiate and discuss the potential usage of blockchain for pharmaceutical supply chain solutions.  \\                                    

\bottomrule

\end{tabularx}
\end{table}

\subsection{Prescription management}
Proper management of prescription is important to ensure the best healthcare service delivery. Misuse of the prescription has been rampant in recent years leading to large-scale problems like Opioid crisis \cite{Skolnick18} . Many blockchain based solutions have been proposed to remove the impediments on proper prescription management.

BlockMedx \cite{BlockMedx18} is using an Ethereum based platform to securely manage prescription processes where all transactions are all securely stored in a blockchain. After a doctor issues a prescription to a patient, the designated pharmacist can verify it via blockchain before issuing the drugs. This system makes the management of controlled drug prescriptions like opioid efficient.  Project Heisenberg \cite{Heisenberg18} is another example of an application using smart contracts on top of Ethereum to track prescriptions. It provides separate portals for patients, doctors, and pharmacies for their stake into the prescription process.
ScriptDrop \cite{ScriptDrop18} works on streamlining pharmacy delivery to patients. They deliver the drugs to the patients, relieving them of the burden of having to show up at the pharmacy for their medications. They also track the usage of medicine (adherence) using virtual assistants. ScriptDrop is using blockchain to track the information about adherence and delivery. ScalaMed \cite{ScalaMed18} offers a blockchain based solution from for medical adherence and tracking of all prescriptions (including historical) around a patient-centric model. It describes the solution as a digital prescription inbox which will solve issues related to medicine mismanagement. Think about the prescription from your general practitioner that you had in the recent past which might be problematic for your current ongoing treatment at a hospital. In the current system, it is challenging to have these different data points in one place that is accessible to the patient and this, in turn, could lead to undesirable incidents. The solution from ScalaMed brings all the information about current and past prescriptions into a central place which helps avoid situations causing cross-reactions of the prescription drugs.

Though several blockchain based solutions are being investigated for prescription management, some centralized conventional system could also provide a solution for certain markets where the involved parties are very few. Close collaboration and joint tool for use by the clinician and pharmacy, where possible, e.g. due to enforcement by regulations, could envision a simple solution like a single sign-on into a common database to have an overview of a patient’s medication.

\subsection{Billing Claims Management}
Financial aspects of medical care are inherently important in the healthcare landscape. This area of financing aspect in healthcare is rife with inefficiencies, mostly related to the trust and transparency, which can potentially be optimized by the use of blockchain.  Blockchain provides a mechanism for direct links between patients (one who makes claims) with the bearers (one who clears the claim), as there is trust inbuilt. Smart contracts can be used in the premium negotiating phases. Data regarding the current health status, medication usage, lifestyle, etc.  tied through blockchain to evolving premiums, through smart contracts. When many parties or intermediaries are involved in the claim handling, there might be a lot of repetitive tasks and checks involved which might be burdensome for the end customer. 

After seeing these opportunities for improvements, some propositions have been brought forth using blockchain for billing claims management and broader financial aspect of care delivery. Gem \cite{Gem18} is using Ethereum to streamline the claim management in healthcare services, among others. It brings the patients, providers, and the insurers together into one ecosystem to provide real-time insights into patient's health journey and ease health claims management. Change healthcare \cite{Change18} is using the HyperLedger Fabric framework \cite{androulaki2018hyperledger} for blockchain based claims and revenue management. HSBlox has unveiled its RevBlox(TM) and CuraBlox(TM) product \cite{HSBlox18} for claims management built on top of their blockchain platform dubbed SETU (Simplified Exchange and Transparency for Users). Besides the trust and openness provided by the blockchain, the company uses machine learning to facilitate an automatic decision on top of claims, for instance, duplicate claim detection or pattern identification of claim denials. Dockchain from Pokidot \cite{Smith17} provides blockchain-based financial data processing in a clinical setting, using features like smart contracts.
Multiple healthcare companies, including insurers like Humana and UnitedHealthcare, have been working on a pilot program \cite{Morse18} using blockchain to maintain and share curated information from healthcare providers. This feature can solve a lot of redundancies and inefficiencies related to insurance claims management. An overview of example initiatives on the use of blockchain enabled solutions for billing and claim management in the healthcare field is outlined in Table \ref{tab:billing_claims}.

\begin{table}[h]
\caption{Companies and initiatives that use/plan to use blockchain based solution for billing and claim management.}
\label{tab:billing_claims}
\begin{tabularx}{\linewidth}{p{0.35\linewidth} | p{0.6\linewidth}}
\toprule
Initiatives & Description \\
\midrule
Gem \cite{Gem18}                                                         & Blockchain based on Ethereum to streamline claim management in healthcare.                                                                                                                                   \\
Change Healthcare \cite{Change18}                                           & Solution based on HyperLedger fabric 1.0 for claims and revenue management.                                                                                                                                  \\
HSBlox \cite{HSBlox18}                                                                    & A blockchain based platform called SETU (Simplified Exchange and Transparency for Users) to provide solutions for claim management.                                                                          \\
Pokitdok \cite{pokidok}                                                                & Provides Dockchain, a blockchain for financial data processing in a clinical setting, using features like smart contracts.                                                                                   \\
Solve.care \cite{SolveCare}                                                                & Blockchain based solution from Solve.care provides decentralized administration of health mostly concerning various health benefits program, preventing misuse and fraud for example.                        \\

HealthNautica \cite{HealthNautica}  collaboration with Factom \cite{factom_whitepaper} & This joint collaboration works on a blockchain solution for claim management and data record management in general.                                                                                          \\

Smartillions \cite{Smartillions}                                                      & The solution from Smartillions used blockchain based system for claims management with payment from an underlying pension fund, with option of all transactions to providers also done with a digital asset. \\
Robomed Network \cite{Robomed}                                                       & Robomed in its blockchain based solution ties payment for a medical procedure to expected clinical outcome, motivating the providers for a first time right medical treatment.                               \\
Quantum Medical Transport collaboration with River Oaks Billing Associates \cite{QuantumMedical}                    & The collaboration is using blockchain based solution for medical billing payments, mostly to make such transactions secure.              
\\
\bottomrule

\end{tabularx}
\end{table}

It is to be noted that while blockchain provides a technological medium for multiple parties to transact with trust, some building blocks would need to be there to facilitate such a transaction. A simple example of this case is the need of standard data formats so that information can be easily exchanged in regard to health conditions, treatments made, and corresponding claims. If a common language is enforced in the underlying blockchain, then the onus is on the multiple parties involved (hospitals, insurers, etc.) to bear the translation overhead from their domain language to the common lingua.

\section{Analytics}
Blockchain can have a huge positive impact on all three pillars of analytics: data, model, and computation.
\
\subsubsection*{Data}  Access to the right data is the key to better analytics.  Data are more valuable than algorithms, and this is especially true in healthcare due to the sensitive nature of healthcare data. With blockchain-enabled HIE, researchers and analytics companies can efficiently acquire the desired data for analysis. Also, confidence in the quality and attribute of the acquired data increases if its provenance is tracked by blockchain. Not only data but the analytics process, for example, training and validation of prediction models, can also be tracked using blockchain. This tracking feature can increase the confidence of healthcare professionals and regulatory entities such as FDA on the prediction models, thereby increasing their probability of being used in the clinical setting.

\subsubsection*{Model} Multiple model predictions can be pooled using a blockchain network to create a final robust model; prediction of an ensemble of models is usually more correct than individual predictions. Also, incentive features such as crypto assets and reputation scores can be used to encourage participation and submission of better models. For example, Numerai \cite{Numerai_whitepaper} is a hedge fund which utilizes crowd-sourced machine learning for financial predictions. A similar mechanism can be set up in health care for use cases such as early detection of cancer. Modelchain \cite{Kuo2018_modelchain} has proposed a framework for crowd-sourced machine learning while preserving patient information; inside a private blockchain network, participating sites share only model parameters which are used to build the final model. In addition to the prediction models, the predictions can come from individuals where correct predictions are rewarded while incorrect ones are penalized. This mechanism can be utilized in public health, for example, to predict the outbreak of epidemic diseases. For example, using Augur \cite{Peterson2018_augur_whitepaper} which is a decentralized prediction platform, a bet has been created to predict the spread of human-borne Nipah virus in United States \cite{Nipah_USA-Augur}. 
\subsubsection*{Computation} With the increase in data quantity and model complexity, the demand for computing resources is rapidly increasing. Research centers and analytics companies either have in-house computing servers or use computing services from Amazon, Google, or Microsoft. With the introduction of the blockchain, decentralized computation has become the third option. Recently there has been few promising projects developing open-source p2p computation platforms such as Golem \cite{Golem_whitepaper}, SONM \cite{SONM_whitepaper}, etc.  They have released some early products and have a consistent user base, although a tiny one.  One of the key challenges in decentralized computation is preserving data privacy while performing computation on them, and it is particularly important in healthcare.  Companies such as Enigma \cite{Engima_whitepaper} and Oasis Labs \cite{Oasis_Labs_arxiv_paper} are developing solutions for privacy-preserving computation on blockchain. Compared to traditional cloud computing services, blockchain promises cheaper computation with greater data security and removes the need of vendor lock-in. However, the blockchain-based solutions are still in the early stages and need to prove their claims in practice. 

\section{Telemedicine} \label{sec:telemedicine}
Telemedicine is another area in healthcare that can benefit from blockchain technology by introducing a trust layer between patients and healthcare professionals. Blockchain-based telemedicine platform can validate professional identity and data integrity, ensure transparency and traceability, and incentivize the players to act fairly by providing incentive metrics such as reputation scores and crypto tokens. Successful development of such platforms may create a global healthcare marketplace in the future effectively balancing the global supply and demand of healthcare resources and needs. In addition, such a network can incorporate the software services such as Artificial Intelligence (AI)  if the AI models are properly validated, regulated, and their performance continuously quantified by some proxy measures.

Within telemedicine, it is likely that the remote diagnostic services will be at the forefront of blockchain adoption. It can be expected that diagnostic services solely based on quantitative and qualitative interpretation of medical data in the absence of a patient will be first to adopt blockchain technology successfully.  A large number of startups in this area corroborates this anticipation. And many of these startups are targeting the services where the diagnosis of a medical condition is based on interpretation of patient-generated imaging data such as dermatology. We found the projects focusing on skin images unusually high in number, most probably due to its close relationship with the beauty industry in addition to the above-mentioned reasons. See Table \ref{tab:telemedicine} for a non-comprehensive list of projects working on blockchain based telemedicine platforms.

   

\begin{table}[]
\centering
\caption{Example initiatives on the use of blockchain for telemedicine}
\label{tab:telemedicine}
\begin{tabularx}{\linewidth}{p{0.2 \linewidth}  | p{0.8\linewidth}}
        \toprule
        \textbf{Comments}   & \textbf{Projects}\\
        \midrule
         & 
        Medicalchain \cite{medicalchain_whitepaper}
        \\ \\

         Dermatolgoy  & 
         OPU Labs\cite{OPU_whitepaper},
         MedX Protocol\cite{medxprotocol_medcredits_website},
         Dermonet \cite{Dermnet_italy_paper}
       \\ \bottomrule
    
    \end{tabularx}
    
\end{table}

\section{Consulting and Blockchain as a service (BaaS)}
Several companies are developing services that can be utilized by existing healthcare organizations to benefit from blockchain. 
Pokidok \cite{pokidok} provides platform as a service to healthcare organizations to integrate blockchain in their existing legacy systems. Their primary services are identity and access management and benefits and claims management. Hashed Health \cite{hashedhealth} provides expertise and services to healthcare enterprises to create new blockchain-based solutions and integrate them with existing systems. Factom \cite{factom_whitepaper} is an enterprise-focused blockchain company that offers solutions such as Harmony Connect and Harmony Integrate; these solutions enable existing applications with blockchain capabilities such as trust and data integrity without the need to deal with blockchain components such as cryptocurrencies, wallets, or network nodes.  IBM \cite{ibm_blockchain_solutions} has introduced IBM Blockchain Platform on top of Hyperledger Fabric \cite{androulaki2018hyperledger} which allows users to create their own blockchain solutions. 
Guardtime provides blockchain based solutions through its KSI technology stack \cite{guardtime_KSI_website}. Recently together with Instant Access Medical and Healthcare Gateway, it launched a blockchain-supported personal care record platform called MyPCR \cite{guardtime_myPCR_website}. 
SSOT Health \cite{ssothealth} is developing a platform to create healthcare blockchains based in Hyperledger.

\section{Challenges} \label{sec:challenges}

Although blockchain holds tremendous potential to improve and add value to the healthcare system and several companies have already started testing it for specific use cases, there are several challenges to be tackled before its mainstream usage. There are various facets to the challenges on the mainstream adoption of the blockchain technology. These facets are for example:
\begin{itemize}
    \item Technical
    \item Legal
    \item Business
    \item Trust issues
\end{itemize}

In this section, we touch upon some of the challenges to be expected for the widespread usage of the blockchain technology in the healthcare sector.

\subsection{Interoperability \& integration with the legacy systems}
Healthcare space has a vast number of technologies, devices, and components, not forgetting the personnel, which all come together to solve the current needs in the space. There are large rooms for improvements, with blockchain technology providing a promise to cover up some of this room. However, the blockchain technology would still be just one piece, though likely very crucial, of the puzzle.  The blockchain technology has to integrate well with existing systems, and the integration is going to be a challenging ordeal because of several reasons such as \textit{interoperability}. That the healthcare space has multiple numbers of devices and device types make it even more difficult.  All relevant parties and stakeholders should all come together to make the use of blockchain technology more pervasive in healthcare.  If blockchain cannot be a catalyst for this cooperation between stakeholders and relevant parties, then it probably does not have much value above and beyond being a simple and useful tool to solve some issues around trust. Many of the other issues around ecosystem building will still be there, regardless of the blockchain. An important aspect of this discussion also relates to the training of healthcare IT personnel. These personnel might be needed to be re-trained on blockchain technology if the blockchain technology would enter into the healthcare space. 

\subsection{Adoption and incentives for participation}
As discussed, adoption of blockchain technology in the healthcare space would require co-ordination and co-operation of multiple stakeholders. These stakeholders, for example, could be the hospital, device manufacturers, care personnel, patients, etc. As the adoption of blockchain technology would require co-operation from multiple stakeholders, with some changes in each of these stakeholder’s current operational and business model, it would be natural for these stakeholders to expect some incentive to participate in the change process. In order to manage these expectations, new business models must be explored that can provide fair incentives to all the stakeholders. The provided incentives should acknowledge the costs and efforts in deploying or adopting a blockchain based technologies, along with any underlying changes in the operational model that the deployed solution entails. Consider the solution for blockchain based patient data management. In this particular scenario, a fair incentive scheme must be devised in a sustainable business model that can be accepted by the patients, device manufactures, backend IT solution providers, hospitals, etc. As this application is still evolving, it would be some time before proven robust incentive models are aplenty. In the meantime, any associated risk of a particular incentive model must be rigorously assessed and quantified as much as possible.

\subsection{Uncertain cost of operation}
While blockchain has promising features such as no need of central authority (and hence no central point of failure), transparency and relatively fast settlement of transactions, the cost of operating blockchain systems are not yet known. Today, a significant amount of resources in healthcare is being spent on personnel, time, and money to build and manage the current traditional information systems and data exchanges. On top of that, there is an overhead of continuously updating the systems, troubleshooting issues, performing backups, worrying about hacks and data breaches. The blockchain-based HIE  system could prove itself to be cost-effective and more efficient compared to the traditional HIE system. This, for example, could come due to the enhanced security by design. Nonetheless, the overall cost factors involved in a blockchain technology-based healthcare services must be robustly assessed in a given business and operational model of a healthcare organization.

\subsection{Regulation}
A solution in the healthcare space needs to adhere to a various set of regulations as the patient’s health and even life is at stake, directly or indirectly. Further, as it concerns health data, the concerns regarding data privacy are also the highest. The blockchain technology being a new technological solution that is being adopted among the early users, one of the challenges remain on how the solutions on top of blockchain technology comply with existing regulations and standards. While existing HIE systems have multiple years to evolve towards meeting the regulatory requirements, the blockchain technology might be still in the evolution phase to find the sweet landing spot within the healthcare space where it can also adequately adhere to set healthcare standards and regulations. To achieve this, a number of pilot deployments and rigorous test and validation of the underlying technological pieces must be considered by the solution providers. The requirements to adhere to regulations could also be an impetus for further technological advancements in the blockchain technology. 

The requirements from recently enforced General Data Protection Rule (GDPR) is one of the cases which is leading discussion and thoughts on how blockchain technology can further evolve to comply with set regulations. Blockchain and GDPR has a dual relationship. On the one hand, GDPR, as an example, enforces provisions like the right to erasure/right to be forgotten, giving a person (data originator) full right and control over his or her data which directly conflicts with the immutable nature of blockchain. Healthcare data are generally stored off-chain and only pointers to the data, usually, its cryptographic hash, is stored in the blockchain. This approach provides a way to make a blockchain application GDPR compliant. However, one can always argue that the cryptographic hash can still be considered a form of the personal data which is a separate discussion in itself.  Over time better trade-offs are bound to come where regulatory compliance feeds very well to only evolution and maturity of blockchain technology for even more widespread adoption. At this point, it is wise to consider incorporating GDPR like regulations in blockchain applications.  On the other hand, from a decentralization perspective, regulations like GDPR and blockchain are perfect partners.  With the increasing awareness among users about their personal data and enforcement of privacy regulations, there has been a surge in the interest around the patient-controlled health care data. Given the existence of user-friendly applications for management of the personal data with proper security and incentive structures, individuals can have access and control to their healthcare data and can participate in p2p global data-exchange required for healthcare research and services, all powered by blockchain.

\subsection{Governance}
The basic premise of the distributed nature of blockchain helps to bring multiple parties into a trusted transaction scheme, without the need of any centralized authorities. However, if we consider how healthcare organizations operate, there can be several different operational models. In certain operational models of blockchain based solutions, it might be imperative to have a certain stakeholder assuming the role of a regulator to govern the overall operation of the blockchain. This governing model might, for example, be needed to meet the regulatory requirements. It is not yet clear how such a governance structure can be managed properly in a system with multiple disparate parties. This aspect of governance will also have a tie-in with the intended incentive schemes. However, as the adoption of blockchain based solutions will evolve in the healthcare space, we would see various solutions also to the requirements of governance. 

\subsection{Scaling}
Not only in healthcare but any industry, the underlying blockchain network has to be scalable for successful application of any blockchain-based solutions. It is likely that, at least in the early phase, several solutions in healthcare will use semi-permissioned blockchains which are scalable and have high transaction throughput at the cost of decentralization. However, there will still be a need for public blockchains for communication among permissioned blockchain networks. Besides, a blockchain-enabled global HIE can only be possible with highly scalable public blockchains.   In current form, public blockchain networks like Bitcoin and Ethereum are not fast and cheap enough to host any decentralized applications on a large scale. However, there have been several ongoing scaling efforts such as lightning network, state channels, plasma chains, sharding, zk-snarks, etc. and some of them are already being adopted in practice. In the future, it is likely that the public blockchains will be fast and cheap enough for their mass adoption.

\section{Conclusion}
Blockchain technology has the potential to solve several problems plaguing the healthcare industry today. As a trust mediator, it can enable novel healthcare solutions; and as an incentive machine, it can enable novel business models that may lead to a new dynamic among various healthcare stakeholders like patients and providers. For example, a patient-centric healthcare model and a global HIE might be realized by virtue of blockchain enabled decentralized trust and incentive structures. Similarly, blockchain based decentralized network/services may minimize vendor lock-in problems in healthcare. 

In this document, we reviewed major use cases of blockchain such as healthcare data management, supply chain management in the pharmaceutical industry, medication adherence, billing/claims management, analytics, etc. Examples of organizations developing blockchain-based applications for these use cases were also presented. The proposed applications range from moonshot projects trying to build a complete decentralized health care ecosystem to specific applications such as data provenance, counterfeit drugs identification, consent management, etc.

Despite the immense potential of blockchain technology and an enormous amount of interest around it, we found that its impact on healthcare is minimal and is still in the early days. Most of the blockchain based healthcare solutions are still in the form of novel concepts represented by whitepapers, prototypes, or only a very small number of working products with a limited user base. However, the field is evolving rapidly; we anticipate a significant positive impact of blockchain in healthcare in the future. Challenges such as interoperability, integration with the existing systems, uncertainty in cost, technological and adoption barrier, regulatory compliance, and scaling have to be successfully tackled to help blockchain make its mark in the healthcare industry.

{\small
\bibliographystyle{unsrtnat}
\bibliography{references}

\begin{thebibliography}{107}
\providecommand{\natexlab}[1]{#1}
\providecommand{\url}[1]{\texttt{#1}}
\expandafter\ifx\csname urlstyle\endcsname\relax
  \providecommand{\doi}[1]{doi: #1}\else
  \providecommand{\doi}{doi: \begingroup \urlstyle{rm}\Url}\fi

\bibitem[OECD(2018)]{health_spending}
OECD.
\newblock Health spending.
\newblock 2018.
\newblock \doi{https://doi.org/https://doi.org/10.1787/8643de7e-en}.
\newblock URL \url{https://www.oecd-ilibrary.org/content/data/8643de7e-en}.

\bibitem[Bell et~al.(2014)Bell, Ohno-Machado, and Grando]{bell2014sharing}
Elizabeth~A Bell, Lucila Ohno-Machado, and M~Adela Grando.
\newblock Sharing my health data: a survey of data sharing preferences of
  healthy individuals.
\newblock In \emph{AMIA Annual Symposium Proceedings}, volume 2014, page 1699.
  American Medical Informatics Association, 2014.

\bibitem[Peter~Behner(2018(accessed Nov 15, 2018))]{pwc_counterfeit}
Dr. Fabian~Wahl Peter~Behner, Dr. Marie-Lyn~Hecht.
\newblock \emph{Fighting counterfeit pharmaceuticals}, 2018(accessed Nov 15,
  2018).
\newblock
  \url{https://www.strategyand.pwc.com/reports/counterfeit-pharmaceuticals}.

\bibitem[Kuo et~al.(2017)Kuo, Kim, and Ohno-Machado]{Kuo2017}
Tsung-Ting Kuo, Hyeon-Eui Kim, and Lucila Ohno-Machado.
\newblock Blockchain distributed ledger technologies for biomedical and health
  care applications.
\newblock \emph{Journal of the American Medical Informatics Association},
  24\penalty0 (6):\penalty0 1211--1220, 2017.
\newblock \doi{10.1093/jamia/ocx068}.
\newblock URL \url{http://dx.doi.org/10.1093/jamia/ocx068}.

\bibitem[Dhillon et~al.(2017)Dhillon, Metcalf, and Hooper]{Dhillon2017}
Vikram Dhillon, David Metcalf, and Max Hooper.
\newblock \emph{Blockchain in Health Care}, pages 125--138.
\newblock Apress, Berkeley, CA, 2017.
\newblock ISBN 978-1-4842-3081-7.
\newblock \doi{10.1007/978-1-4842-3081-7_9}.
\newblock URL \url{https://doi.org/10.1007/978-1-4842-3081-7_9}.

\bibitem[Engelhardt and Espinosa(2017)]{Engelhardt2017}
Mark~A Engelhardt and Diego Espinosa.
\newblock {Hitching Healthcare to the Chain : An Introduction to Blockchain
  Technology in the Healthcare Sector An Introduction to Blockchain Technology
  in the Healthcare Sector}.
\newblock 7\penalty0 (10):\penalty0 22--35, 2017.

\bibitem[Randall et~al.(2017)Randall, Goel, and Abujamra]{Randall2017}
David Randall, Pradeep Goel, and Ramzi Abujamra.
\newblock {Blockchain Applications and Use Cases in Health Information
  Technology}.
\newblock \emph{Journal of Health {\&} Medical Informatics}, 8\penalty0
  (3):\penalty0 8--11, 2017.
\newblock \doi{10.4172/2157-7420.1000276}.

\bibitem[Sharma(2018(accessed October 18, 2018))]{Sharma2018_FCCCO}
Richa Sharma.
\newblock \emph{Blockchain in Healthcare}, 2018(accessed October 18, 2018).
\newblock
  \url{http://www.fccco.org/uploads/publications/Blockchaininhealthcare_FCCCO_RS.pdf}.

\bibitem[Nakamoto(2008)]{nakamoto2008bitcoin}
Satoshi Nakamoto.
\newblock Bitcoin: A peer-to-peer electronic cash system.
\newblock 2008.

\bibitem[Rogaway and Shrimpton(2004)]{rogaway2004cryptographic}
Phillip Rogaway and Thomas Shrimpton.
\newblock Cryptographic hash-function basics: Definitions, implications, and
  separations for preimage resistance, second-preimage resistance, and
  collision resistance.
\newblock In \emph{International workshop on fast software encryption}, pages
  371--388. Springer, 2004.

\bibitem[Otte et~al.(2017)Otte, de~Vos, and Pouwelse]{otte2017trustchain}
Pim Otte, Martijn de~Vos, and Johan Pouwelse.
\newblock Trustchain: A sybil-resistant scalable blockchain.
\newblock \emph{Future Generation Computer Systems}, 2017.

\bibitem[LeMahieu(2018(accessed November 10,
  2018){\natexlab{a}})]{nano_whitepaper}
Colin LeMahieu.
\newblock \emph{Nano: A Feeless Distributed Cryptocurrency Network},
  2018(accessed November 10, 2018){\natexlab{a}}.
\newblock \url {https://nano.org/en/whitepaper}.

\bibitem[Saleh(2018)]{saleh2018blockchain}
Fahad Saleh.
\newblock Blockchain without waste: Proof-of-stake.
\newblock 2018.

\bibitem[bit(2018 (accessed November 15, 2018))]{bitcoinscript}
\emph{Script - Bitcoin Wiki}, 2018 (accessed November 15, 2018).
\newblock \url{https://en.bitcoin.it/wiki/Script}.

\bibitem[Wood(2014)]{wood2014ethereum}
Gavin Wood.
\newblock Ethereum: A secure decentralised generalised transaction ledger.
\newblock \emph{Ethereum project yellow paper}, 151:\penalty0 1--32, 2014.

\bibitem[Peterson et~al.(2018(accessed October 18, 2018))Peterson, Krug, Zoltu,
  Williams, and Alexander]{Peterson2018_augur_whitepaper}
Jack Peterson, Joseph Krug, Micah Zoltu, Austin Williams, and Stephanie
  Alexander.
\newblock \emph{Augur: a Decentralized Oracle and Prediction Market Platform},
  2018(accessed October 18, 2018).
\newblock \url{https://www.augur.net/whitepaper.pdf}.

\bibitem[zrx(2018 (accessed October, 2018))]{zrx_whitepaper}
2018 (accessed October, 2018).
\newblock URL \url{https://0xproject.com/pdfs/0x_white_paper.pdf}.
\newblock 0x: An open protocol for decentralized exchange on the Ethereum
  blockchain.

\bibitem[LeMahieu(2018(accessed November 10, 2018){\natexlab{b}})]{nem}
Colin LeMahieu.
\newblock \emph{NEM - Distributed Ledger Technology}, 2018(accessed November
  10, 2018){\natexlab{b}}.
\newblock \url {https://nem.io/}.

\bibitem[Androulaki et~al.(2018)Androulaki, Barger, Bortnikov, Cachin,
  Christidis, De~Caro, Enyeart, Ferris, Laventman, Manevich,
  et~al.]{androulaki2018hyperledger}
Elli Androulaki, Artem Barger, Vita Bortnikov, Christian Cachin, Konstantinos
  Christidis, Angelo De~Caro, David Enyeart, Christopher Ferris, Gennady
  Laventman, Yacov Manevich, et~al.
\newblock Hyperledger fabric: a distributed operating system for permissioned
  blockchains.
\newblock In \emph{Proceedings of the Thirteenth EuroSys Conference}, page~30.
  ACM, 2018.

\bibitem[Iris(2018 (accessed Nov15, 2018))]{stellar}
Iris.
\newblock \emph{Develop the world's new financial system}, 2018 (accessed
  Nov15, 2018).
\newblock \url{https://www.stellar.org/}.

\bibitem[Neo-project()]{neo-project}
Neo-project.
\newblock \emph{NEO}.
\newblock \url{https://neo.org/}.

\bibitem[Zheng et~al.(2017)Zheng, Xie, Dai, Chen, and Wang]{zheng2017overview}
Zibin Zheng, Shaoan Xie, Hongning Dai, Xiangping Chen, and Huaimin Wang.
\newblock An overview of blockchain technology: Architecture, consensus, and
  future trends.
\newblock In \emph{Big Data (BigData Congress), 2017 IEEE International
  Congress on}, pages 557--564. IEEE, 2017.

\bibitem[Wav(2018(accessed Nov 15, 2018))]{Waves_whitepaper}
Waves whitepaper, 2018(accessed Nov 15, 2018).
\newblock \url{https://blog.wavesplatform.com/waves-whitepaper-164dd6ca6a23}.

\bibitem[Duffield(2018(accessed Nov 15, 2018))]{dash_whitepaper}
Diaz Duffield.
\newblock \emph{Dash: A Payments-Focused Cryptocurrency}, 2018(accessed Nov 15,
  2018).
\newblock \url{https://github.com/dashpay/dash/wiki/Whitepaper}.

\bibitem[bts(2018(accessed Nov 15, 2018))]{bts_whitepaper}
\emph{The BitShares Blockchain}, 2018(accessed Nov 15, 2018).
\newblock
  \url{https://www.bitshares.foundation/papers/BitSharesBlockchain.pdf}.

\bibitem[Brown(2017)]{brown2017introducing}
Richard~Gendal Brown.
\newblock Introducing r3 corda: A distributed ledger designed for financial
  services, 2016, 2017.

\bibitem[Ene(2018(accessed Nov 15, 2018))]{EnergyWebF}
Energy web foundation, 2018(accessed Nov 15, 2018).
\newblock \url{http://energyweb.org}.

\bibitem[B3i(2018(accessed Nov 15, 2018))]{B3iInsurance}
B3i insurance, 2018(accessed Nov 15, 2018).
\newblock \url{https://b3i.tech/}.

\bibitem[Mon(2018(accessed Nov 15, 2018))]{MonaxBlockchain}
Monax - blockchain explainer, 2018(accessed Nov 15, 2018).
\newblock \url{https://monax.io/learn/blockchains}.

\bibitem[Saw(2018(accessed Nov 15, 2018))]{Sawtooth}
Hyperledger sawtooth, 2018(accessed Nov 15, 2018).
\newblock \url{https://sawtooth.hyperledger.org}.

\bibitem[Goldreich and Oren(1994)]{goldreich1994definitions}
Oded Goldreich and Yair Oren.
\newblock Definitions and properties of zero-knowledge proof systems.
\newblock \emph{Journal of Cryptology}, 7\penalty0 (1):\penalty0 1--32, 1994.

\bibitem[Rowley(2018)]{rowley_2018}
Jason Rowley.
\newblock With at least \$1.3 billion invested globally in 2018, vc funding for
  blockchain blows past 2017 totals, May 2018.
\newblock URL
  \url{https://techcrunch.com/2018/05/20/with-at-least-1-3-billion-invested-globally-in-2018-vc-funding-for-blockchain-blows-past-2017-totals/}.

\bibitem[Harris-Braun et~al.(2018 (accessed November 01, 2018))Harris-Braun,
  Luck, and Brock]{holochain_whitepaper}
Eric Harris-Braun, Nickolas Luck, and Arthur Brock.
\newblock \emph{Holochain Scalabel Agen-centric Distributed Computing}, 2018
  (accessed November 01, 2018).
\newblock
  \url{https://github.com/holochain/holochain-proto/blob/whitepaper/holochain.pdf}.

\bibitem[Baird et~al.(2018 (accessed November 01, 2018))Baird, Harmon, and
  Madsen]{Hydera-hashgraph_whitepaper}
Leemon Baird, Mance Harmon, and Paul Madsen.
\newblock \emph{Hedera: A Governing Council \& Public Hashgraph Network}, 2018
  (accessed November 01, 2018).
\newblock \url{https://www.hedera.com/}.

\bibitem[Ekblaw and Azaria(2017)]{Ekblaw2017MedRec}
Ariel Ekblaw and Asaf Azaria.
\newblock {MedRec: Medical Data Management on the Blockchain}.
\newblock \emph{Viral Communications}, 12 2017.
\newblock URL \url{https://viral.media.mit.edu/pub/medrec}.
\newblock https://viral.media.mit.edu/pub/medrec.

\bibitem[Mcfarlane et~al.(2017)Mcfarlane, Beer, Brown, and
  Prendergast]{Mcfarlane2017}
Chrissa Mcfarlane, Michael Beer, Jesse Brown, and Nelson Prendergast.
\newblock {Patientory - Whitepaper}.
\newblock \penalty0 (May):\penalty0 1--19, 2017.

\bibitem[Web(2018 (accessed October 18, 2018))]{Philips_TNW_Oct2018}
The~Next Web.
\newblock \emph{Philips will challenge tech giants to bring blockchain to
  healthcare}, 2018 (accessed October 18, 2018).
\newblock
  \url{https://thenextweb.com/blockchain/2018/10/17/philips-solve-healthcare-data-breaches-with-blockchain/}.

\bibitem[Xia et~al.(2017)Xia, Sifah, Asamoah, Gao, Du, and Guizani]{Xia2017}
Qi~Xia, Emmanuel~Boateng Sifah, Kwame~Omono Asamoah, Jianbin Gao, Xiaojiang Du,
  and Mohsen Guizani.
\newblock {MeDShare: Trust-Less Medical Data Sharing among Cloud Service
  Providers via Blockchain}.
\newblock \emph{IEEE Access}, 5:\penalty0 14757--14767, 2017.
\newblock ISSN 21693536.
\newblock \doi{10.1109/ACCESS.2017.2730843}.

\bibitem[Iry(2018(accessed October 18, 2018))]{Iryo_whitepaper}
\emph{IRYO Network Technical Whitepaper}, 2018(accessed October 18, 2018).
\newblock \url{https://iryo.io/iryo_whitepaper.pdf}.

\bibitem[ope(2018(accessed November 6, 2018))]{openEHR_website}
\emph{openEHR}, 2018(accessed November 6, 2018).
\newblock \url{www.openehr.org}.

\bibitem[gem(2018(accessed November 6, 2018))]{gem_health}
\emph{Gem Health}, 2018(accessed November 6, 2018).
\newblock \url{https://enterprise.gem.co/health/ }.

\bibitem[Zhang et~al.(2018)Zhang, White, Schmidt, Lenz, and
  Rosenbloom]{Zhang2018_FHIRChain}
Peng Zhang, Jules White, Douglas~C Schmidt, Gunther Lenz, and S~Trent
  Rosenbloom.
\newblock {FHIRChain : Applying Blockchain to Securely and Scalably Share
  Clinical Data}.
\newblock \emph{Computational and Structural Biotechnology Journal},
  16:\penalty0 267--278, 2018.
\newblock ISSN 2001-0370.
\newblock \doi{10.1016/j.csbj.2018.07.004}.
\newblock URL \url{https://doi.org/10.1016/j.csbj.2018.07.004}.

\bibitem[Roehrs et~al.(2017)Roehrs, Andr{\'{e}}, and Righi]{Roehrs2017_OMNIphr}
Alex Roehrs, Cristiano Andr{\'{e}}, and Rosa Righi.
\newblock {OmniPHR : A distributed architecture model to integrate personal
  health records}.
\newblock \emph{Journal of Biomedical Informatics}, 71:\penalty0 70--81, 2017.
\newblock ISSN 1532-0464.
\newblock \doi{10.1016/j.jbi.2017.05.012}.
\newblock URL \url{http://dx.doi.org/10.1016/j.jbi.2017.05.012}.

\bibitem[med(2018(accessed November 6, 2018))]{medicalchain_whitepaper}
\emph{Medicalchain Whitepaper 2.1}, 2018(accessed November 6, 2018).
\newblock \url{https://medicalchain.com/Medicalchain-Whitepaper-EN.pdf }.

\bibitem[doc(2018(accessed November 6, 2018))]{doc.ai_website}
2018(accessed November 6, 2018).
\newblock \url{ https://doc.ai/ }.

\bibitem[hea(2018(accessed November 6, 2018))]{hearthy.co_whitepaper}
\emph{Hearthy Foundation}, 2018(accessed November 6, 2018).
\newblock \url{http://hearthy.co/assets/images/Hearthy-whitepaper.pdf}.

\bibitem[fac(2018(accessed November 7,
  2018){\natexlab{a}})]{factom_gates_healthcare}
2018(accessed November 7, 2018){\natexlab{a}}.
\newblock
  \url{https://bitcoinmagazine.com/articles/gates-foundation-grant-boosts-factom-s-blockchain-based-medical-record-development-1479492383/}.

\bibitem[enc(2018(accessed Nov 5, 2018))]{encrypgen_website}
\emph{Encrypgen}, 2018(accessed Nov 5, 2018).
\newblock \url{https://encrypgen.com/}.

\bibitem[Grishin et~al.(2018(accessed November 6, 2018))Grishin, Obbad, Estep,
  Cifric, Zhao, and Church]{nebula_genomics_whitepaper}
Dennis Grishin, Kamal Obbad, Preston Estep, Mirza Cifric, Yining Zhao, and
  George Church.
\newblock \emph{Nebula Genomics: Blockchain-enabled genomic data sharing and
  analysis platform}, 2018(accessed November 6, 2018).
\newblock \url{https://www.nebula.org/assets/Nebula_Genomics_Whitepaper.pdf}.

\bibitem[lun(2018(accessed Nov 5, 2018))]{luna_dna_website}
\emph{lunaDNA}, 2018(accessed Nov 5, 2018).
\newblock \url{https://lunadna.com/ }.

\bibitem[Kulemin et~al.(2018(accessed Nov 5, 2018))Kulemin, Popov, and
  Gorbachev]{zenome_whitepaper}
Nikolay Kulemin, Sergey Popov, and Alexey Gorbachev.
\newblock \emph{The Zenome Project: Whitepaper blockchain-based genomic
  ecosystem}, 2018(accessed Nov 5, 2018).
\newblock \url{https://zenome.io/download/whitepaper.pdf }.

\bibitem[Hahnel(2018(accessed Nov 5, 2018))]{hahnel2018_genomes_whitepaper}
Mark Hahnel.
\newblock \emph{The Genomes.io Lightpaper- Blockchain enabled genome security
  from the moment it is sequenced}, 2018(accessed Nov 5, 2018).
\newblock \url{https://genomes.io/assets/uploads/Genomes-whitepaper.pdf }.

\bibitem[shi(2018(accessed Nov 5, 2018))]{shivom_website}
\emph{Shivom}, 2018(accessed Nov 5, 2018).
\newblock \url{https://shivom.io//}.

\bibitem[Ortega et~al.(2018(accessed November 6, 2018))Ortega, Oikonomou, Ding,
  Rees-Lee, Alexandria, and Jensen]{Ortega2018_ETDB-Caltech}
Davi~R Ortega, Catherine~M. Oikonomou, H.~Jane Ding, Prudence Rees-Lee,
  Alexandria, and Grant~J Jensen.
\newblock \emph{ETDB-Caltech: a blockchain-based distributed public database
  for electron tomography}, 2018(accessed November 6, 2018).
\newblock \url{https://www.biorxiv.org/content/early/2018/10/25/453662}.

\bibitem[Patel(2018)]{Patel2018_image_exchange}
Vishal Patel.
\newblock A framework for secure and decentralized sharing of medical imaging
  data via blockchain consensus.
\newblock \emph{Health Informatics Journal}, 0\penalty0 (0):\penalty0
  1460458218769699, 2018.
\newblock \doi{10.1177/1460458218769699}.
\newblock URL \url{https://doi.org/10.1177/1460458218769699}.
\newblock PMID: 29692204.

\bibitem[Labs(2018(accessed November 7, 2018))]{OPU_whitepaper}
OPU Labs.
\newblock \emph{OPU whitepaper 2018}, 2018(accessed November 7, 2018).
\newblock
  \url{https://ico.opu.ai/wp-content/uploads/2018/04/Opucoin_Whitepaper_v2.0.pdf?x38818
  }.

\bibitem[Protocol(2018(accessed November 7,
  2018))]{medxprotocol_medcredits_website}
MedX Protocol.
\newblock 2018(accessed November 7, 2018).
\newblock \url{https://medxprotocol.com }.

\bibitem[Der(2018(accessed November 7, 2018))]{Dermnet_italy_paper}
\emph{A Blockchain Approach Applied to a Teledermatology Platform in the
  Sardinian Region (Italy)}, 2018(accessed November 7, 2018).
\newblock \url{http://www.mdpi.com/2078-2489/9/2/44/pdf }.

\bibitem[aki(2018(accessed November 6, 2018))]{akiri}
\emph{Akiri Switch}, 2018(accessed November 6, 2018).
\newblock \url{https://akiri.com }.

\bibitem[Enc(2018(accessed October 10, 2018))]{Encrypgen_whitepaper2017}
\emph{The Clinical and Investment Potential in the Gene-Chain Project The
  Unprecedented Growth of Genomic Data}, 2018(accessed October 10, 2018).
\newblock
  \url{https://icotimeline.com/wp-content/uploads/2017/07/Gene-Chain-Whitepaper.pdf}.

\bibitem[Lab(2018(accessed September, 2018))]{Labii2018}
2018(accessed September, 2018).
\newblock
  \url{https://blog.labii.com/2018/03/using-blockchain-technology-inelectronic-lab-notebook-eln.html}.

\bibitem[Ber(2018 (accessed September, 2018))]{Berstein18}
2018 (accessed September, 2018).
\newblock URL \url{https://www.bernstein.io/}.
\newblock Bernstein - Blockchain for Intellectual Property.

\bibitem[iPl(2018 (accessed September, 2018))]{iPlexus18}
2018 (accessed September, 2018).
\newblock URL
  \url{https://www.innoplexus.com/life-science-ai-products-solutions-3-2/iplexus}.
\newblock iPlexus.

\bibitem[iee(2018(accessed November 6,
  2018))]{ieee_blockchain_for_clinical_trials}
2018(accessed November 6, 2018).
\newblock \url{ https://blockchain.ieee.org/standards/clinicaltrials }.

\bibitem[Worley(2018)]{worley18}
R.~R. Worley.
\newblock Scrybe: A blockchain ledger for clinical trials.
\newblock In \emph{IEEE Clinical Trials Forum}, 2018.

\bibitem[Ravaud(2017)]{Ravaud17}
M.~B. Ravaud.
\newblock Blockchain technology for improving clinical research quality.
\newblock \emph{Trials}, 2017.

\bibitem[Blo(2017 (accessed September 2018))]{BlockRx17}
2017 (accessed September 2018).
\newblock URL \url{https://www.blockrx.com/white-paper/}.
\newblock Retrieved from BlockRx: The Pharmaceutical Blockchain of Value:.

\bibitem[Med(2017 (accessed September 2018))]{MediLedger2017}
2017 (accessed September 2018).
\newblock URL \url{https://www.mediledger.com/}.
\newblock Retrieved from Building an Open Network for the Pharmaceutical Supply
  Chain:.

\bibitem[Craib et~al.(2018(accessed October 29, 2018))Craib, Bradway, and
  Dunn]{ambrosus_whitepaper}
Richard Craib, Geo Bradway, and Xander Dunn.
\newblock \emph{Ambrosus White paper}, 2018(accessed October 29, 2018).
\newblock \url {https://ambrosus.com/assets/en/Ambrosus-White-Paper.pdf}.

\bibitem[Mod(2018 (accessed September,2018))]{Modum18}
2018 (accessed September,2018).
\newblock URL \url{https://modum.io/solution/products}.
\newblock Retrieved from MODSense T1:.

\bibitem[Blo(2018 (accessed September, 2018))]{Blockverify18}
2018 (accessed September, 2018).
\newblock URL \url{http://www.blockverify.io/}.
\newblock Retrieved from Blockchain Based Anti-Counterfeit Solution:.

\bibitem[DHL(2018(accessed September, 2018))]{DHLAccenture}
2018(accessed September, 2018).
\newblock
  \url{https://newsroom.accenture.com/news/dhl-and-accenture-unlock-the-power-of-blockchain-in-logistics.htm}.

\bibitem[Imp(2018(accessed September, 2018))]{ImperialOneNetwork}
2018(accessed September, 2018).
\newblock
  \url{https://www.onenetwork.com/2012/01/one-network-enterprises-and-imperial-logistics-partner-to-achieve-supply-chain-excellence-visibility-and-flexibility-creates-next-generation-supply-chain/}.

\bibitem[Aut(2018(accessed September, 2018))]{Authentag}
2018(accessed September, 2018).
\newblock \url{https://www.authentag.com/ }.

\bibitem[Eas(2018(accessed September, 2018))]{EasySight}
2018(accessed September, 2018).
\newblock
  \url{https://www.prnewswire.com/news-releases/ibm-and-hejia-launch-blockchain-based-supply-chain-financial-services-platform-for-pharmaceutical-procurement-300437935.html}.

\bibitem[GFT(2018(accessed September, 2018))]{GFT}
2018(accessed September, 2018).
\newblock \url{https://www.gft.com }.

\bibitem[Myt(2018(accessed September, 2018))]{Mytigate}
2018(accessed September, 2018).
\newblock \url{http://mytigate.com/ }.

\bibitem[IEE(2018(accessed September, 2018))]{IEEEPharma}
2018(accessed September, 2018).
\newblock \url{https://blockchain.ieee.org/standards/pharmasupply}.

\bibitem[Skolnick(2018)]{Skolnick18}
P.~Skolnick.
\newblock The opoid epidemic: Crisis and solutions.
\newblock \emph{Annual Review of Pharmacology and Toxicology}, 2018.

\bibitem[Blo(2018 (accessed September 2018))]{BlockMedx18}
2018 (accessed September 2018).
\newblock URL \url{https://blockmedx.com/en/}.
\newblock Retrieved from BlockMedx - Secure e-prescribing Platform:.

\bibitem[Heisenberg(2018 (accessed September 2018))]{Heisenberg18}
P.~Heisenberg, 2018 (accessed September 2018).
\newblock URL \url{https://github.com/tylerdiaz/Heisenberg}.
\newblock Retrieved from Solving prescription/pharmaceutical logistics using
  smart contracts:.

\bibitem[Scr(2018 (accessed September 2018))]{ScriptDrop18}
2018 (accessed September 2018).
\newblock URL \url{https://www.scriptdrop.co/}.
\newblock Retrieved from Prescription Delivery In Workflow:.

\bibitem[Sca(2018 (accessed September 2018))]{ScalaMed18}
2018 (accessed September 2018).
\newblock URL \url{https://www.scalamed.com/}.
\newblock Retrieved from ScalaMed:.

\bibitem[Gem(2018 (accessed September 2018))]{Gem18}
2018 (accessed September 2018).
\newblock URL \url{https://enterprise.gem.co/health/}.
\newblock Retrieved from Gem - Health:.

\bibitem[Healthcare(2018 (accessed September 2018))]{Change18}
C.~Healthcare.
\newblock Retrieved from better healthcare, 2018 (accessed September 2018).
\newblock URL \url{https://www.changehealthcare.com/}.
\newblock Improved Lives:.

\bibitem[HSB(2018 (accessed September 2018))]{HSBlox18}
2018 (accessed September 2018).
\newblock URL \url{https://hsblox.com/solutions/}.
\newblock Retrieved from HSBlox:.

\bibitem[Smith(2017)]{Smith17}
W.~B. Smith.
\newblock DOKCHAIN: INTELLIGENT AUTOMATION IN HEALTHCARE, 2017.
\newblock URL
  \url{https://pokitdok.com/wp-content/themes/pokitdok2017/dokchain/static/data/DokChainWhitepaper20170926Draft.pdf}.

\bibitem[Morse()]{Morse18}
S.~(2018 Morse.
\newblock April). healthcare it news.
\newblock URL
  \url{https://www.healthcareitnews.com/news/optum-unitedhealthcare-humana-others-launch-blockchain-pilot}.
\newblock Retrieved from Optum, UnitedHealthcare, Humana, others launch
  blockchain pilot:.

\bibitem[pok(2018(accessed November 7, 2018))]{pokidok}
2018(accessed November 7, 2018).
\newblock \url{https://pokitdok.com/ }.

\bibitem[Sol(2018(accessed September, 2018))]{SolveCare}
2018(accessed September, 2018).
\newblock \url{https://solve.care/}.

\bibitem[Hea(2018(accessed September, 2018))]{HealthNautica}
2018(accessed September, 2018).
\newblock \url{http://www.healthnautica.com}.

\bibitem[fac(2018(accessed November 7, 2018){\natexlab{b}})]{factom_whitepaper}
2018(accessed November 7, 2018){\natexlab{b}}.
\newblock
  \url{https://www.factom.com/wp-content/uploads/2018/10/Factom_Whitepaper_v1.2.pdf}.

\bibitem[Sma(2018(accessed September, 2018))]{Smartillions}
2018(accessed September, 2018).
\newblock \url{http://www.smartillions.ch/ }.

\bibitem[Rob(2018(accessed September, 2018))]{Robomed}
2018(accessed September, 2018).
\newblock \url{https://www.robomed.io/}.

\bibitem[Qua(2018(accessed September, 2018))]{QuantumMedical}
2018(accessed September, 2018).
\newblock
  \url{https://globenewswire.com/news-release/2018/04/04/1460245/0/en/Quantum-Medical-Transport-Inc-Enters-Joint-Venture-Partnership-With-River-Oaks-Billing-Associates-for-Medical-Billing-Blockchain.html}.

\bibitem[Craib et~al.(2018(accessed October 18, 2018))Craib, Bradway, and
  Dunn]{Numerai_whitepaper}
Richard Craib, Geo Bradway, and Xander Dunn.
\newblock \emph{Numeraire: A Cryptographic Token for Coordinating Machine
  Intelligence and Preventing Overfitting}, 2018(accessed October 18, 2018).
\newblock \url{https://numer.ai/static/media/whitepaper.29bf5a91.pdf}.

\bibitem[Kuo and Ohno-Machado(2018(accessed October 18,
  2018))]{Kuo2018_modelchain}
Tsung-Ting Kuo and Lucila Ohno-Machado.
\newblock \emph{ModelChain: Decentralized Privacy-Preserving Healthcare
  Predictive Modeling Framework on Private Blockchain Networks}, 2018(accessed
  October 18, 2018).
\newblock \url{https://arxiv.org/pdf/1802.01746.pdf}.

\bibitem[Health(2018(accessed October 18, 2018))]{Nipah_USA-Augur}
Liquidity Health.
\newblock \emph{Human Borne Nipah Virus (NiV) to be confirmed in the USA
  on/before midnight UTC 30 Jun 2019? (Twitter @LiquidityHealth)},
  2018(accessed October 18, 2018).
\newblock
  \url{https://predictions.global/augur-markets/human-borne-nipah-virus-niv-to-be-confirmed-in-the-usa-on-before-midnight-utc-30-jun-2019-twitter-liquidityhealth-0x9fadd5b149174e762e92a7805dfa2099b21d348d}.

\bibitem[Gol(2018(accessed October 18, 2018))]{Golem_whitepaper}
\emph{The Golem Project Crowdfunding Whitepaper}, 2018(accessed October 18,
  2018).
\newblock \url{https://golem.network/crowdfunding/Golemwhitepaper.pdf}.

\bibitem[SON(2018(accessed October 18, 2018))]{SONM_whitepaper}
\emph{SONM Decentralized Fog Computing Platform}, 2018(accessed October 18,
  2018).
\newblock \url{ https://docs.sonm.com/}.

\bibitem[Zyskind and Nathan(2018(accessed October 18,
  2018))]{Engima_whitepaper}
Guy Zyskind and Alex~’Sandy’ Nathan, Oz~Pentland.
\newblock \emph{Enigma: Decentralized Computation Platform with Guaranteed
  Privacy}, 2018(accessed October 18, 2018).
\newblock \url{ https://enigma.co/enigma_full.pdf}.

\bibitem[Cheng et~al.(2018(accessed October 18, 2018))Cheng, Zhang, Kos, Hynes,
  Johnson, Juels, and Miller]{Oasis_Labs_arxiv_paper}
Raymond Cheng, Fan Zhang, Warren Kos, Jernej~He, Nicholas Hynes, Noah Johnson,
  Ari Juels, and Dawn Miller, Andrew~andSong.
\newblock \emph{Ekiden : A Platform for Confidentiality-Preserving ,
  Trustworthy , and Performant Smart Contracts}, 2018(accessed October 18,
  2018).
\newblock \url{ https://arxiv.org/pdf/1804.05141.pdf}.

\bibitem[has(2018(accessed November 7, 2018))]{hashedhealth}
2018(accessed November 7, 2018).
\newblock \url{ https://hashedhealth.com/ }.

\bibitem[ibm(2018(accessed November 7, 2018))]{ibm_blockchain_solutions}
2018(accessed November 7, 2018).
\newblock \url{ https://www.ibm.com/blockchain/solutions }.

\bibitem[gua(2018(accessed November 7,
  2018){\natexlab{a}})]{guardtime_KSI_website}
2018(accessed November 7, 2018){\natexlab{a}}.
\newblock \url{ https://guardtime.com/technology }.

\bibitem[gua(2018(accessed November 7,
  2018){\natexlab{b}})]{guardtime_myPCR_website}
2018(accessed November 7, 2018){\natexlab{b}}.
\newblock \url{
  https://guardtime.com/blog/world-s-first-blockchain-supported-personal-care-record-platform-launched-by-guardtime-and-partners
  }.

\bibitem[sso(2018(accessed November 7, 2018))]{ssothealth}
2018(accessed November 7, 2018).
\newblock \url{ https://www.ssothealth.org/ }.

\end{thebibliography}
}
\end{document}